\documentclass[12pt]{article}
\usepackage{amsmath}
\usepackage{graphicx,psfrag,epsf}
\usepackage{enumerate}
\usepackage{natbib}
\usepackage{url} % not crucial - just used below for the URL 
\usepackage[english]{babel}
\usepackage{amsmath}
\usepackage{graphicx,psfrag,epsf}
\usepackage{enumerate}
\usepackage{natbib}
\usepackage{amsfonts}
\usepackage[latin1]{inputenc}
\usepackage[ulem=normalem]{changes}
\usepackage{cancel}
\usepackage{paralist}
\usepackage{multirow}
\usepackage[small]{caption}
\usepackage{setspace}
\usepackage{url}
\usepackage{bbm}
\usepackage{dsfont}
\usepackage{sectsty}
\usepackage{lipsum}
\usepackage{epstopdf}
\usepackage{mathtools}

\usepackage{amssymb}
\usepackage{mathrsfs}
\usepackage{bm}

\newcommand{\blind}{1}

\begin{document}

\def\spacingset#1{\renewcommand{\baselinestretch}%
{#1}\small\normalsize} \spacingset{1}

%%%%%%%%%%%%%%%%%%%%%%%%%%%%%%%%%%%%%%%%%%%%%%%%%%%%%%%%%%%%%%%%%%%%%%%%%%%%%%

\if1\blind
{
  \title{\bf A notion of depth for sparse functional data}
  \author{
	Carlo Sguera\thanks{UC3M-Santander Big Data Institute, Universidad Carlos III de Madrid, Getafe, Spain. E-		mail: carlo.sguera@uc3m.es.}
	\hspace{0.05cm} and Sara L{\'o}pez-Pintado\thanks{Department of Health Sciences, Northeastern University, 		Boston, USA. E-mail: s.lopez-pintado@northeastern.edu. Partial support from the National Institute of 			Mental Health (grant number: 1R21MH120534-01) is acknowledged.}
    }
  \date{}
  \maketitle 
} \fi

\if0\blind
{
  \bigskip
  \bigskip
  \bigskip
  \begin{center}
    {\LARGE\bf Title}
\end{center}
  \medskip
} \fi

\bigskip
\begin{abstract}
Data depth is a well-known and useful nonparametric tool for analyzing functional data. It provides a novel way of ranking a sample of curves from the center outwards and defining robust statistics, such as the median or trimmed means. It has also been used as a building block for functional outlier detection methods and classification. Several notions of depth for functional data were introduced in the literature in the last few decades. These functional depths can only be directly applied to samples of curves measured on a fine and common grid. In practice, this is not always the case, and curves are often observed at sparse and subject dependent grids. In these scenarios the usual approach consists in estimating the trajectories on a common dense grid, and using the estimates in the depth analysis. This approach ignores the uncertainty associated with the curves estimation step. Our goal is to extend the notion of depth so that it takes into account this uncertainty. Using both functional estimates and their associated confidence intervals, we propose a new method that allows the curve estimation uncertainty to be incorporated into the depth analysis. We describe the new approach using the modified band depth although any other functional depth could be used. The performance of the proposed methodology is illustrated using simulated curves in different settings where we control the degree of sparsity. Also a real data set consisting of female medflies egg-laying trajectories is considered. The results show the benefits of using uncertainty when computing depth for sparse functional data.
\end{abstract}

\noindent%

\spacingset{1.45} % DON'T change the spacing!

\section{Introduction and Motivation}
Functional data analysis is an exciting developing area in statistics where the basic unit of observation is a function/curve. Many different statistical methods, such as principal components, analysis of variance, and linear regression, have been extended to functional data. In the last two decades there has been an intensive development of different notions of data depth which have been proven to be a powerful nonparametric tool for analyzing functional data. In general, a data depth is a function that measures the centrality (or outlyingness) of an observation within a population or sample. It provides a novel way of ranking observations from the center outwards and allows the definition of robust statistics such as medians, trimmed means and central regions for functional data. Moreover, data depth is often used as a building block for developing classification and outlier-detection techniques.  Several notions of depth for functional data have been introduced in the literature (e.g., \cite{fraiman2001trimmed}, \cite{cuevas2007robust}, \cite{cuesta2008random}, \cite{lopez2009concept}, \cite{lopez2011half} or \cite{sguera2014spatial}). Functional depths have been originally proposed for sample of curves that are measured on a common and dense grid. In practice, curves are often observed at subject-dependent and/or sparse grids. The main approach in the literature for dealing with this situation is based on estimating the individual trajectories on an artificially chosen common dense grid of points and using these estimated curves as observed data in a depth analysis (e.g., \cite{lopez2011depth}). Up to now, a functional depth analysis usually ignores the inherent uncertainty associated with the preliminary curve estimation step. In this paper we propose a general approach for calculating the depth of sparsely observed functions and we take uncertainty into account by analyzing with a depth function both functional estimates and their associated confidence intervals.

We present the new approach using: (1) the modified band depth ($MBD$, \cite{lopez2009concept}) as functional depth; (2) the iterated expectation and variance method ($IEV$, \cite{goldsmith2013corrected}) to obtain estimates and confidence intervals. Note that the proposed methodology can be generalized to any functional depth and to any curve estimation method that provides confidence intervals. The paper is arranged as follows: in Section \ref{sec:depth} we give a general overview of the notion of functional depth focusing on $MBD$. In Section \ref{sec:mbdu} we propose the novel approach to compute the depth of sparse functional data taking into account the uncertainty in the estimation and define a new version of $MBD$ that we call ``modified band depth under uncertainty'' ($MBD_{U}$). Section \ref{sec:sstudy} shows the performance of $MBD_{U}$ in a simulation study where curves with different degrees of sparsity are generated. In Section \ref{sec:medfly} we use $MBD_{U}$ in a real data example consisting of female medflies egg-laying trajectories (\cite{carey1998relationship}).

\section{The modified band depth}
\label{sec:depth}
Given a probability distribution, a statistical depth assigns to each point a real non-negative bounded value that measures the centrality of the point with respect to its distribution. Several depth definitions for multivariate data have been proposed and analyzed by \cite{mahalanobis1936generalized}, \cite{tukey1975mathematics}, \cite{oja1983descriptive}, \cite{liu1990notion}, \cite{liu1993quality}, \cite{chaudhuri1996geometric}, \cite{koshevoy1997zonoid}, \cite{liu1999multivariate}, \cite{rousseeuw1999regression}, \cite{vardi2000multivariate} and \cite{zuo2003projection} among others. \cite{liu1990notion} and \cite{zuo2000general} introduced and studied key properties a depth should satisfy. However, most of these depths are computationally intractable and not well defined in high-dimensions and functional spaces. 

In the last two decades several notions of depth have been proposed for functional data (see, e.g., \cite{fraiman2001trimmed}; \cite{lopez2007depth}, \cite{cuesta2008random}, \cite{lopez2009concept}, \cite{lopez2011half}; \cite{lopez2007functional}, \cite{cuevas2007robust}; \cite{gervini2012outlier}, \cite{sguera2014spatial}, \cite{chakraborty2014data}, \cite{narisetty2016extremal}, among others). Functional depths provide a novel way of ranking functions from the center outwards and robust location estimators such as median or trimmed means can be defined using depth functions. Moreover, those curves from the sample with low depth can be considered as potential outliers and depth-based outlier detection rules for functional data have been recently introduced in the literature (see, e.g., \cite{hubert2015multivariate}, \cite{arribas2014shape} and \cite{sun2011functional}, \cite{dai2017outlyingness}, \cite{sguera2016functional}, \cite{azcorra2018unsupervised}, among others). Also, robust nonparametric tests for functional data based on functional depths have been proposed in the literature (see, e.g., \cite{lopez2009concept}, \cite{sun2012functional}, \cite{lopez2017robust}, \cite{flores2018homogeneity}). The notion of depth can also be used for developing nonparametric classification methods (see e.g., \cite{jornsten2004clustering}, \cite{li2012dd}, \cite{sguera2014spatial} and \cite{cuesta2017hbox}).

In this paper we focus on the modified band depth for functional data introduced in \cite{lopez2009concept}, which is based on the graphic representation of the curves. It satisfies desirable theoretical properties and is computationally fast. It provides a natural and novel way of ordering curves from the center outwards, and can be used to generalize classical order statistics to functional data.

Let $(\mathcal{C}(I),\mathcal{P})$ be the space of continuous real valued functions on the compact interval $I \in \mathbb{R}$ with the supremum norm $\|\cdot\|_{\infty}$ and the probability measure $\mathcal{P}$. The modified band depth of a function $y$ in $(\mathcal{C}(I),\mathcal{P})$  is defined as $MBD(y,\mathcal{P})=E_{Y_1,Y_2}[\lambda(A(y;Y_{1},Y_{2}))]$, where $\lambda=\frac{\lambda_{L}}{\lambda_{L} (I)}$, $\lambda_{L}$ is Lebesgue measure in $\mathbb{R}$ and

\begin{equation}
\label{eq:A}
A\left(y; Y_{1}, Y_{2}\right) = \left\{s\in I: \min_{j=1,2} Y_{j}(s) \leq y(s) \leq \max_{j=1,2} Y_{j}(s)\right\}.
\end{equation}

\noindent Basically, $MBD(y,\mathcal{P})$ measures how long the curve $y$ is expected to be inside a stochastic band determined by two random functions $Y_{1}$ and $Y_{2}$ from $(\mathcal{C},\mathcal{P})$. By Fubini's theorem one can express $MBD$ as

\begin{equation}
MBD(y; \mathcal{P})=\int_{\cal I} SD(y(s); \mathcal{P}_{Y(s)}) ds,
\label{MBD2}
\end{equation}

\noindent where 

\begin{equation*}
SD(y(s); \mathcal{P}_{Y(s)}) = P(\min({Y_{1}(s),Y_{2}(s)}) \leq y(s) \leq \max({Y_{1}(s),Y_{2}(s)}))
\end{equation*}

\noindent is the standard univariate simplicial depth of $y$ at location $s$. $MBD$ can be computed in a very fast and efficient way using the algorithm in \cite{sun2012functional}.

Let $Y_{i}(s)$, with $1 \leq i \leq n$, be a sample of $n$ functions from $(\mathcal{C}(I),\mathcal{P})$, and denote this sample as $\bm{Y} = \left\{Y_{i}(s)\right\}_{i=1}^{n}$. The sample modified band depth of a given curve $Y_{i}$ with respect to the whole functional data set $\bm{Y}$ is given by

\begin{equation}
\label{eq:MBD}
MBD(Y_{i}; \bm{Y}) = {{n}\choose{2}}^{-1} \sum_{1 \leq i_{1} \leq i_{2} \leq n} \lambda \left(A\left(Y_{i}; Y_{i_{1}}, Y_{i_{2}}\right)\right).
\end{equation}

Intuitively, the sample modified band depth ($MBD(Y_{i}; \bm{Y})$) measures in average for how long the function $Y_{i}$ is contained in the band determined by any two functions $Y_{i_{1}},Y_{i_{2}}$ from the sample.  It can be seen as a measure of centrality/similarity between $Y_i$ and the sample curves. The modified band depth satisfies natural and desirable depth properties such as: non-degenarcy, invariance, maximality at center, decreasing with respect to the deepest point, semi-continuity, and consistency. See Table 1 in \cite{gijbels2017general} for details about these properties and comparisons with other depth notions. \cite{mosler2012general} and \cite{nieto2016topologically} provide more discussion on functional data depth properties.  

To apply the notion of depth to a sample of functions, the curves have to be evaluated at the same regular grid. In practice, this is rarely the case, since many times the functions are observed at different sparse points, and therefore a preliminary step to estimate the sample functions in a common fine grid is needed. In the next section we propose an approach to calculate the depth of sparse functions taking into account the uncertainty in the estimation of the curves, and we implement it using $MBD$. Note that our approach could be applied to any notion of functional depth.

\section{A new modified band depth under uncertainty}
\label{sec:mbdu}
Although no official definition exists of ``dense'' or ``sparse'' functional data, generally, if the number of observed points per curve is larger than some order of the sample size, $n$, then the functional data are referred to as ``dense'' (see \cite{zhang2016sparse} for a detailed discussion on this topic). Nevertheless, even if the data is dense, some preliminary step is usually needed to have the sample defined on the same common grid of points when the curves are observed at subject specific grid or to denoise the observed sample. As observed by \cite{yao2005functional}, individual smoothing of the curves when data are sparse in general does not work well, and therefore the curves need to be estimated borrowing information from other curves. They proposed a method to reconstruct functional observations from sparse data which relies on functional principal components (FPC) analysis. \cite{goldsmith2013corrected} modified the method proposed by \cite{yao2005functional} defining a bootstrap-based version which accounts for uncertainty in the FPC decomposition and is known as iterated expectation and variance method. IEV is the method we use to reconstruct functional data and obtain confidence intervals of such estimates. All the details about IEV are reported in the appendix.

\begin{figure}[!htbp]
\centering
\includegraphics[width=0.66\textwidth]{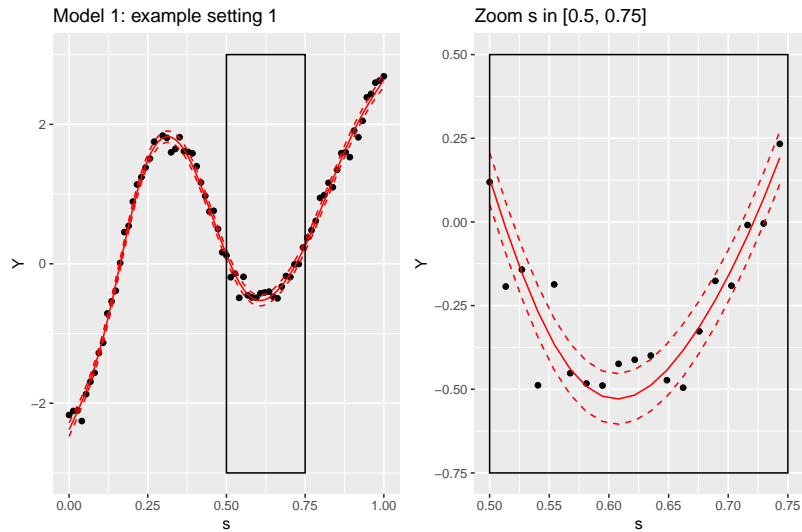}
\captionsetup{width=0.75\textwidth}
\caption{Example of densely observed functional data evaluated at black points together with its $IEV$ estimate (solid line) and 95\% confidence interval (dashed lines) (left). Same example zoomed in and showing only a proportion of the domain (right).}
\label{fig:f02}
\end{figure}

\begin{figure}[!htbp]
\centering
\includegraphics[width=0.66\textwidth]{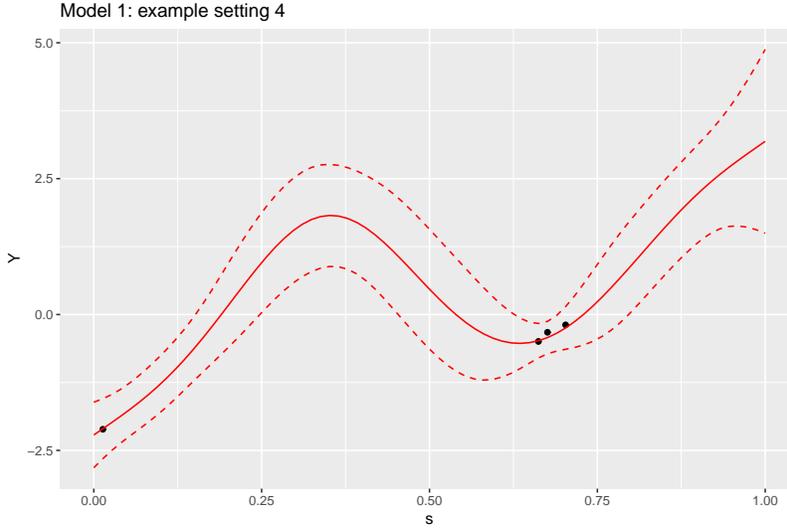}
\captionsetup{width=0.75\textwidth}
\caption{Example of sparsely observed functional data evaluated at black points together with its $IEV$ estimate (solid line) and 95\% confidence interval (dashed lines).}
\label{fig:f03}
\end{figure}

To show the differences between dealing with densely or sparsely observed functional data, see Figures \ref{fig:f02} and \ref{fig:f03}.

To obtain Figures \ref{fig:f02} and \ref{fig:f03} we have considered a simulation model (Model 1) and two simulation settings (settings 1 and 4, respectively) that we present in detail in Section \ref{sec:sstudy}. In Figure \ref{fig:f02} we represent a densely observed curve together with its associated IEV estimate and 95\% confidence intervals. In Figure \ref{fig:f03} we represent a sparsely observed curve together with its associated IEV estimate and 95\% confidence intervals. In the dense case there is little estimation uncertainty and the confidence interval is so narrow that it is hard to appreciate. For this reason we also zoom in on a portion of the domain. In the sparse case there is a lot of estimation uncertainty and the confidence interval is very wide. 

Figures \ref{fig:f02} and \ref{fig:f03} illustrate the differences between dense and sparse functional data. Our next goal is to present a new depth that does not depend only on the estimated curves but also on their associated confidence intervals.

Let $\tilde{\bm{Y}}$ be a functional data set consisting of functions that are observed with measurement error and on different subject specific finite grids which could be irregularly spaced. We use $IEV$ to obtain curves estimates and confidence intervals for $\tilde{\bm{Y}}$. Let $\hat{\bm{Y}}$ be the functional data set containing the curves estimates provided by $IEV$ when applied to $\tilde{\bm{Y}}$. Similarly, let $\hat{\bm{Y}}_{ub}$ and $\hat{\bm{Y}}_{lb}$ be the functional data sets composed by the confidence intervals upper and lower bounds, respectively. Note that estimates and confidence intervals are evaluated on a common and dense grid denoted as $\bm{s}_{g}$.

When performing a data depth analysis on $\tilde{\bm{Y}}$, a standard approach consists on simply applying any depth function, for example $MBD$, to $\hat{\bm{Y}}$. We propose to take into account that $\hat{\bm{Y}}$ is the result of a preliminary estimation step and incorporate in the depth analysis its related uncertainty. In particular, we propose to include the information contained in $\hat{\bm{Y}}_{ub}$ and $\hat{\bm{Y}}_{lb}$ in the following way: first, create the auxiliary enlarged functional data set of size $3n$, $\hat{\bm{Y}}_{U}$, defined as the union of $\hat{\bm{Y}}_{ub}, \hat{\bm{Y}}$ and $\hat{\bm{Y}}_{lb}$; second, compute $MBD$ for $\hat{\bm{Y}}_{U}$; third, assign the following depth value to each $\hat{Y}_{i}(\bm{s}_{g}) = \hat{Y}_{i}$:

\begin{equation}
\label{eq:MBDU}
MBD_{U}\left(\hat{Y}_{i}; \hat{\bm{Y}}\right) = 
\frac{MBD\left(\hat{Y}_{ub,i}; \hat{\bm{Y}}_{U}\right) + MBD\left(\hat{Y}_{i}; \hat{\bm{Y}}_{U}\right) + MBD\left(\hat{Y}_{lb,i}; \hat{\bm{Y}}_{U}\right)}{3},
\end{equation}

\noindent where $\hat{Y}_{lb,i} = \hat{Y}_{lb,i}(\bm{s}_{g})$ and $\hat{Y}_{ub,i} = \hat{Y}_{ub,i}(\bm{s}_{g})$. Note that $MBD_{U}\left(\hat{Y}_{i}; \hat{\bm{Y}}\right)$ is defined as the average depth with respect to $\hat{\bm{Y}}_{U}$ of three functions: the estimated curve, the confidence interval upper bound and the confidence interval lower bound. By using these elements, $MBD_{U}$ is taking into account the estimation uncertainty, and therefore our proposal represents an alternative to the standard approach, $MBD\left(\hat{Y}_{i}; \hat{\bm{Y}}\right)$, which only depends on the curves estimates. 

So far we are omitting that $MBD_{U}$ depends on a tuning parameter $\alpha \in (0, 1)$ which controls the width of the $100(1-\alpha)\%$ confidence intervals. Although, we have shown that for a given $\alpha$, the confidence interval is wider when the curve is more sparsely observed. Moreover, in a preliminary study we have observed that for a given functional sample:

\begin{itemize}
\item low values of $\alpha$ generate very wide intervals and versions of $MBD_{U}$ that give too much weight to estimation uncertainty.

\item high values of $\alpha$ generate too narrow intervals and versions of $MBD_{U}$ that resemble excessively to $MBD$.
\end{itemize}

Let $MBD_{U, \alpha}\left(\hat{Y}_{i}; \hat{\bm{Y}}\right)$ be the version of \eqref{eq:MBDU} that we obtain using a given $\alpha$. Seeking for a version of $MBD_{U}$ that systematically employs intermediate values of $\alpha$, we have defined the following procedure to set $\alpha$:

\begin{enumerate}
\item Given $\tilde{\bm{Y}}$ and estimated $\hat{\bm{Y}}$ using $IEV$, compute $MBD_{U, \alpha}\left(\hat{Y}_{i}; \hat{\bm{Y}}\right)$ for $\alpha \in \linebreak \left\{0.05, 0.06, \ldots, 0.98, 0.99\right\}$ and each observation ($1 \leq i \leq n$). Moreover, compute $MBD\left(\hat{Y}_{i}; \hat{\bm{Y}}\right)$ for each observation. Let $MBD_{U, \alpha}(\hat{\bm{Y}})$ be the sample depth values obtained using our proposal and $MBD(\hat{\bm{Y}})$ be the sample depth values obtained using $MBD$ applied only to the curves estimates.
	
\item For each $\alpha \in \left\{0.05, 0.06, \ldots, 0.98, 0.99\right\}$, compute the Spearman rank correlation coefficient $\rho_{S}$ between $MBD(\hat{\bm{Y}})$ and $MBD_{U, \alpha}(\hat{\bm{Y}})$, i.e., $\rho_{S}(\hat{\bm{Y}}; \alpha) = \rho_{S}(MBD(\hat{\bm{Y}}), MBD_{U, \alpha}(\hat{\bm{Y}}))$. In what follows we omit the subscript $S$ to refer to the Spearman rank correlation coefficient, hence $\rho_S = \rho$. 
	
\item Let $\alpha^{*}$ be the largest $\alpha$ such that  

\begin{equation}
\label{eq:rho95}
\rho(\hat{\bm{Y}}; \alpha) \leq 0.95. 
\end{equation}

If $\alpha^{*}$ exists, use $\alpha^{*}$ and compute $MBD_{U, \alpha^{*}}(\hat{\bm{Y}})$. If $\alpha^{*}$ does not exist, do not compute $MBD_{U, \alpha}(\hat{\bm{Y}})$.
\end{enumerate}

Note that the larger $\alpha$ the narrower the confidence intervals and the closer is $MBD_U$ to $MBD$. This procedure explores different versions of $MBD_{U}$, from versions that use wide confidence intervals (low values of $\alpha$) and give more importance to estimation uncertainty, to versions that use narrow confidence intervals (high values of $\alpha$). Note that $\alpha^{*}$ is chosen in a data-driven way, looking at the relationship between $MBD(\hat{\bm{Y}})$ and $MBD_{U, \alpha}(\hat{\bm{Y}})$. The relationship is analyzed using Spearman rank correlation coefficients because the ranking is certainly one of the most valuable outputs on any depth analysis. This analysis is done searching for a value of $\alpha$ such that $MBD(\hat{\bm{Y}})$ and $MBD_{U, \alpha}(\hat{\bm{Y}})$ differ between them, but not excessively (for this reason we set the threshold for $\rho$ to 0.95). If $\alpha^{*}$ exists, we argue that $MBD_{U, \alpha^{*}}(\hat{\bm{Y}})$ might be preferable to $MBD(\hat{\bm{Y}})$ since it is taking estimation uncertainty into account. If $\alpha^{*}$ does not exist, there is no need to use $MBD_{U, \alpha}(\hat{\bm{Y}})$ instead of $MBD(\hat{\bm{Y}})$ since they never differ excessively. Note that the computational cost to set $\alpha^{*}$ is negligible with respect to the computational cost of obtaining the bootstrap-based $IEV$ estimates, which is a step required by both the standard approach, $MBD(\hat{\bm{Y}})$, and $MBD_{U, \alpha}(\hat{\bm{Y}})$.

In the next section we use an extensive simulation study to compare the behavior of $MBD$ and $MBD_{U}$.

\section{Simulation study}
\label{sec:sstudy}
In this section we present the results of a simulation study designed to evaluate the performance of $MBD_{U}$ in different settings. We consider the models described in \cite{zhang2016sparse} to generate functional data. Four settings with different degrees of sparsity are considered for generating the number of observed evaluation points, $J_{i}$, $i = 1, \ldots, n$:

\begin{itemize}
\item Setting 1: $J_{i}$ are i.i.d. from a discrete uniform distribution on the set $\left\{\lfloor n/8 \rfloor, \lfloor (n+1)/8 \rfloor, \ldots, \lfloor 3n/8 \rfloor\right\}$, where $\lfloor x \rfloor$ indicates the integer part of $x$.
	
\item Setting 2: $J_{i} = n/4$ or i.i.d from a discrete uniform distribution on the set $\left\{2, 3, 4, 5\right\}$ with equal probability.		 
	
\item Setting 3: $J_{i} = n/4$ or i.i.d from a discrete uniform distribution on the set $\left\{2, 3, 4, 5\right\}$ with probability equal to $n^{-1/4}$ and $1-n^{-1/4}$, respectively.		 
	
\item Setting 4: $J_{i}$ are i.i.d. from a discrete uniform distribution on the set $\left\{2, 3, 4, 5\right\}$.
\end{itemize}

\noindent Setting 1 provides dense data while setting 4 provides sparse data. Settings 2 and 3 provide dense and sparse data with the difference that under setting 2 approximately half observations are dense and half are sparse while under setting 3 the majority of observations are sparse. Nevertheless, in all settings curves are observed on grids that usually differ from one curve to another, and therefore it is always necessary to estimate them on a common grid. We use $IEV$ to obtain the estimates, and the number of bootstrap iterations that we use is 100.

Following the ideas of \cite{zhang2016sparse}, we use as model 1 a model with true mean and covariance functions defined as follows:

\begin{equation}
\label{eq:mu&cov}
\mu(s) = \frac{3}{2} \sin\left(3\pi\left(s+\frac{1}{2}\right)\right) + 2s^{3}, \qquad \Sigma\left(s, s'\right) = \sum_{k=1}^{4} \lambda_{k}\phi_{k}(s)\phi_{k}\left(s'\right), \quad s, s' \in [0,1],
\end{equation}

\noindent where $\lambda_{k} = 1/(k+1)^2, k = 1, \ldots, 4$ and

\begin{align*}
\phi_{1}(s) =& \sqrt{2}\cos(2\pi s),\, \phi_{2}(s) = \sqrt{2}\sin(2\pi s),\\
\phi_{3}(s) =& \sqrt{2}\cos(4\pi s),\, \phi_{4}(s) = \sqrt{2}\sin(4\pi s).
\end{align*}

\noindent Then, functional data are generated using

\begin{equation}
\label{eq:y}
\tilde{Y}_{i}(s) = Y_{i}(s) + \epsilon_{i}(s) = \mu(s) + \sum_{k=1}^{4} \xi_{ik} \phi_{k}(s) + \epsilon_{i}(s),\quad 1 \leq i \leq n,  
\end{equation}

\noindent where $\xi_{ik}$ are i.i.d from $N(0, \lambda_{k})$ and $\epsilon_{i}(s)$ are i.i.d from $N(0, 0.01)$, and independent between them.

\begin{figure}[!htbp]
\centering
\includegraphics[width=0.66\textwidth]{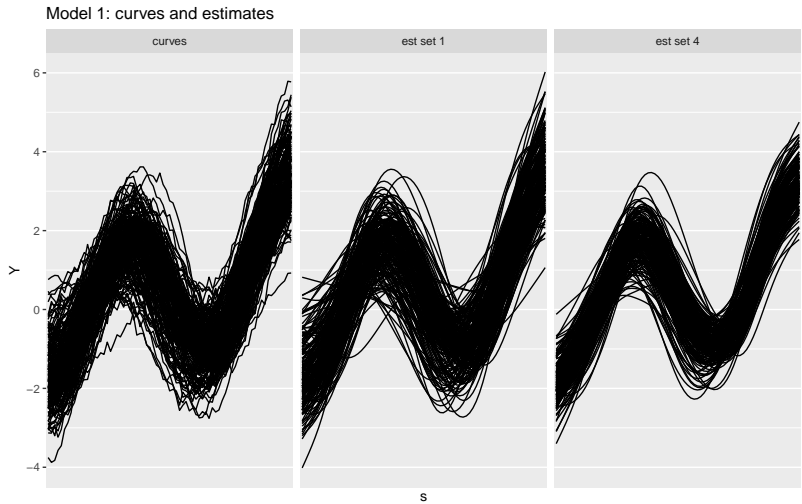}
\captionsetup{width=0.75\textwidth}
\caption{Functional data set generated by model 1 (left). $IEV$ estimates of the curves generated from model 1 after applying setting 1 (center). $IEV$ estimates of the curves generated from model 1 after applying setting 4 (right).}
\label{fig:f01}
\end{figure}

We use model 1 in the following way: first, we generate $n=200$ functional data evaluated at $\lfloor 3n/8 \rfloor = 75$ equidistant points in $[0,1]$. Then, we apply settings from 1 to 4 to obtain dense and sparse scenarios. Once a setting is applied, we use $IEV$ to reconstruct the curves and build the confidence intervals. In Figure \ref{fig:f01} we report an example of a functional data set before applying any setting (left), its IEV estimates after applying setting 1 (center) and its IEV estimates after applying setting 4 (right).

Observing Figure \ref{fig:f01} it is possible to appreciate some minor differences between the curves estimates under setting 1 (Figure \ref{fig:f01}, center) and under setting 4 (Figure \ref{fig:f01}, right). Note that we appreciate more difference between these two very different sparse setting if we consider the IEV confidence intervals (Figures \ref{fig:f02} and \ref{fig:f03}) instead of just the estimates (Figure \ref{fig:f01}).

\begin{figure}[!htbp]
\centering
\includegraphics[width=0.66\textwidth]{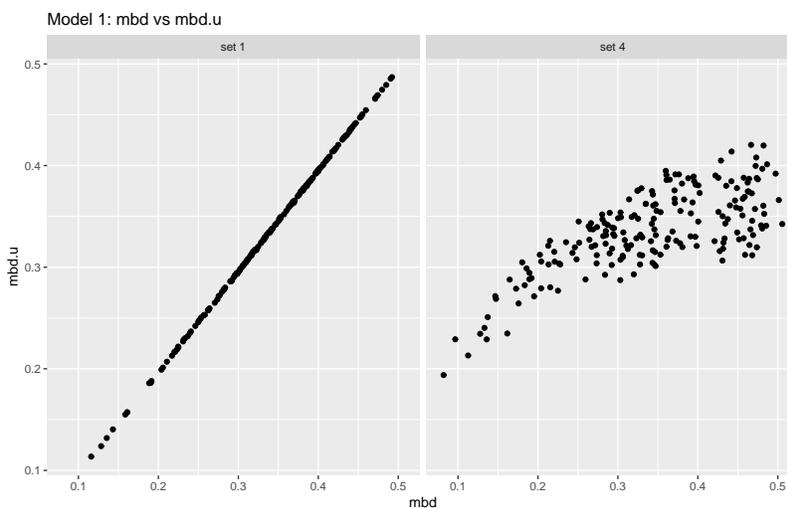}
\captionsetup{width=0.75\textwidth}
\caption{Scatter plots of $(MBD(\hat{\bm{Y}}), MBD_{U, 0.05}(\hat{\bm{Y}}))$ obtained after implementing setting 1 (left) and setting 4 (right) to a functional data set generated from model 1.}
\label{fig:f04}
\end{figure}

To illustrate the differences between $MBD$ and $MBD_{U}$, we have computed both depth measures for the estimated curves in Figure \ref{fig:f01} under setting 1 (center panel) and setting 4 (right panel). For this example we use $\alpha = 0.05$ to obtain standard 95\% confidence intervals. Figure \ref{fig:f04} shows the scatter plots of ``$MBD$ versus $MBD_{U}$'' under setting 1 and 4 in the left and right panel, respectively.

From Figure \ref{fig:f04} it can be seen that when data are densely observed in practice there is no difference between $MBD$ and $MBD_{U}$ since the degree of uncertainty is low (Figure \ref{fig:f04}, left). However, when data are sparsely observed the degree of uncertainty is higher and we observe differences between $MBD$ and $MBD_{U}$ (Figure \ref{fig:f04}, right). 

\begin{figure}[!htbp]
\centering
\includegraphics[width=0.66\textwidth]{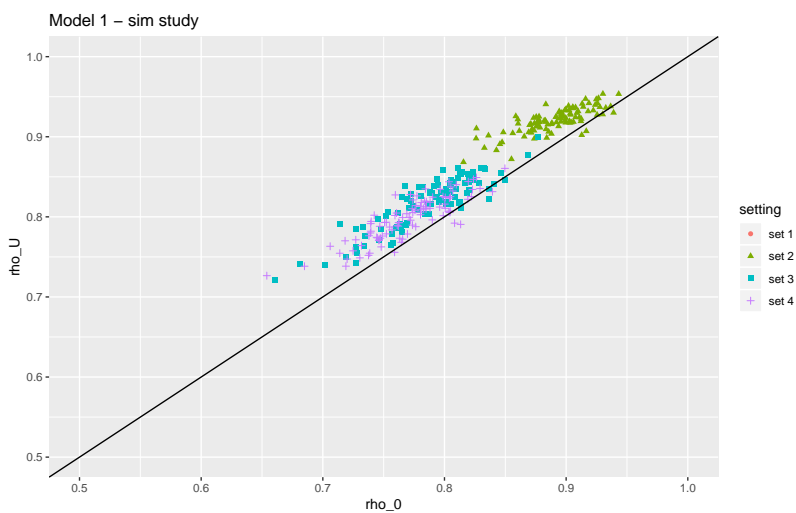}
\captionsetup{width=0.75\textwidth}
\caption{Model 1: scatter plot of $(\rho_{0}, \rho_{U})$ under settings 1-4.}
\label{fig:f05}
\end{figure}

To further explore these differences and analyze which depth approach performs better, we use a simulation study in which we simulate 100 data sets using model 1 and apply each one of the settings. For each data set generated by model 1, say $\bm{\tilde{Y}}$, and before implementing settings 1-4, we compute the ``true'' depth $MBD(\bm{Y})$, i.e., the depth values obtained using $MBD$ and $\bm{Y}$, the functional data set without measurement errors $\epsilon_{i}(s)$. Then, after generating $\bm{\tilde{Y}}$ under each setting and using $IEV$, we compute $MBD(\bm{\hat{Y}})$ and $MBD_{U, \alpha^{*}}(\bm{\hat{Y}})$, and their respective Spearman rank correlation coefficients with respect to the benchmark $MBD(\bm{Y})$, that is, $\rho_{0} = \rho(MBD(\bm{Y}), MBD(\hat{\bm{Y}}))$ and $\rho_{U} = \rho(MBD(\bm{Y}), MBD_{U, \alpha^{*}}(\hat{\bm{Y}}))$. Recall that $\alpha^{*}$ is set using the data-driven procedure described in Section \ref{sec:mbdu}. Therefore, for each pair model-setting, we observe 100 values of $\rho_{0}$ and $\rho_{U}$, and also $\alpha^{*}$. In Figure \ref{fig:f05} we report all the pairs $(\rho_{0}, \rho_{U})$. The different types of points indicate the different settings. 

In Figure \ref{fig:f05} there are no points associated to setting 1, and this is due to the fact that according to the procedure to set $\alpha$ presented in Section \ref{sec:mbdu} there is no $\alpha$ such that \eqref{eq:rho95} holds, and therefore $MBD$ and $MBD_U$ provide very similar rankings in this dense setting and there is no need to use $MBD_U$. For the rest of the settings almost all the points are above the diagonal, indicating a consistent better performance of $MBD_U$ with respect to $MBD$.

To provide more information, in Table \ref{tab:t01} we report the median values of $\rho_{0}$, $\rho_{U}$, the quantity $\Delta_{\rho} = \frac{\rho_{U}-\rho_{0}}{\rho_{0}}\times 100\%$ and $\alpha^{*}$.

\begin{table}[!htbp]
\captionsetup{width=0.75\textwidth}
\centering
\scalebox{1}{
\begin{tabular}{c|cccc}
\hline
& \multicolumn{4}{c}{median values}\\
\hline
& $\rho_{0}$ & $\rho_{U}$ & $\Delta_{\rho}$ & $\alpha^{*}$\\
\hline
setting 1 & 1.00 & - & - & -\\
setting 2 & 0.89 & 0.92 & 3.34\% & 0.23\\
setting 3 & 0.79 & 0.82 & 3.83\% & 0.30\\
setting 4 & 0.77 & 0.80 & 3.61\% & 0.29\\
\hline
\end{tabular}}
\caption{Model 1: median values of $\rho_{0}$, $\rho_{U}$, $\Delta_{\rho}$ and $\alpha^{*}$ under settings 1-4.}
\label{tab:t01}
\end{table}

Based on Figure \ref{fig:f05} and Table \ref{tab:t01} we first notice something expected: both $MBD(\bm{\hat{Y}})$ and $MBD_{U, \alpha^{*}}(\bm{\hat{Y}})$ have a stronger relationship with the benchmark $MBD(\bm{Y})$ under setting 2, for which the median values of $\rho_{0}$ and $\rho_{U}$ are 0.89 and 0.92, respectively. Under setting 3 the median values drop to 0.79 for $\rho_{0}$ and 0.82 for $\rho_{U}$, while under setting 4 they slightly decrease to 0.77 and 0.80, respectively. Moreover, these values and Figure \ref{fig:f05} show that the overall performances of $MBD_{U, \alpha^{*}}(\bm{\hat{Y}})$ are better than the ones of $MBD(\bm{\hat{Y}})$, no matter the setting, which means that in general there is a stronger relationship between the benchmark $MBD(\bm{Y})$ and $MBD_{U, \alpha^{*}}(\bm{\hat{Y}})$ than with $MBD(\bm{\hat{Y}})$. Indeed, if for each data set we look at $\Delta_{\rho}$, its median values are 3.34\% (setting 2), 3.83\% (setting 3) and 3.61\% (setting 4). With respect to $\alpha^{*}$, recall that under setting 1 $\alpha^{*}$ never exists, which means that $\rho(MBD(\hat{\bm{Y}}), MBD_{U, \alpha}(\hat{\bm{Y}}))$ is always greater than 0.95, while its median values are 0.23, 0.30 and 0.29 under settings 2, 3 and 4, respectively. To interpret this pattern, recall that, for a given $\alpha$, $IEV$ confidence intervals are naturally wider in sparse settings (see Figures \ref{fig:f02} and \ref{fig:f03}). Therefore, the fact that the median value of $\alpha^{*}$ is lower in setting 2 than in setting 3 or 4 means that our data-driven procedure is more conservative under sparse scenarios, and it selects lower confidence levels.

Besides model 1, we consider three additional models:

\begin{itemize}
\item Model 2 is obtained modifying the distribution of $\xi_{ik}$ in \eqref{eq:y}: the scores $\xi_{ik}$ are drawn with equal probability from either $N(-\sqrt{\frac{\lambda_{k}}{2}}, \lambda_{k}/2)$ or $N(\sqrt{\frac{\lambda_{k}}{2}}, \lambda_{k}/2)$. Model 2 is based on a model considered by \cite{goldsmith2013corrected}.

\item Model 3 is obtained modifying the set of $\lambda_{k}$ in \eqref{eq:y}: $\lambda_{k} = 1/(k+1), k = 1, \ldots, 4$. Model 3 generates functional data using a truncated Karhunen-Lo\`eve expansion with more balanced components.

\item Model 4 is obtained modifying the distribution of $\epsilon_{i}(s)$ in \eqref{eq:y}: $\epsilon_{i}(s)$ are i.i.d from $N(0, 0.02)$. Model 4 generates functional data that are observed with more noise than under model 1.
\end{itemize}

The results for models 2, 3 and 4 are reported in this section in Table \ref{tab:t02} and in the appendix in Figures \ref{fig:f06}, \ref{fig:f07} and \ref{fig:f08}. 

\begin{table}[!htbp]
\captionsetup{width=0.75\textwidth}
\centering
\scalebox{1}{
\begin{tabular}{c|c|cccc}
\hline
& & \multicolumn{4}{c}{median values}\\
\hline
& & $\rho_{0}$ & $\rho_{U}$ & $\Delta_{\rho}$ & $\alpha^{*}$\\
\hline
\multirow{4}{*}{model 2} & setting 1 & 1.00 & - & - & -\\
& setting 2 & 0.88 & 0.91 & 4.19\% & 0.23\\
& setting 3 & 0.76 & 0.80 & 4.39\% & 0.28\\
& setting 4 & 0.75 & 0.78 & 3.77\% & 0.29\\
\hline
\multirow{4}{*}{model 3} & setting 1 & 1.00 & - & - & -\\
& setting 2 & 0.87 & 0.92 & 5.41\% & 0.27\\
& setting 3 & 0.76 & 0.80 & 6.06\% & 0.33\\
& setting 4 & 0.73 & 0.77 & 5.74\% & 0.36\\
\hline
\multirow{4}{*}{model 4} & setting 1 & 0.99 & - & - & -\\
& setting 2 & 0.87 & 0.91 & 4.00\% & 0.24\\
& setting 3 & 0.76 & 0.80 & 4.22\% & 0.30\\
& setting 4 & 0.74 & 0.77 & 3.84\% & 0.31\\
\hline
\end{tabular}}
\caption{Model 2, 3 and 4: median values of $\rho_{0}$, $\rho_{U}$, $\Delta_{\rho}$ and $\alpha^{*}$ under settings 1-4.}
\label{tab:t02}
\end{table}

Here we report the main findings on these additional models:

\begin{itemize}
\item Under setting 1 and models 2, 3 and 4, the procedure to set $\alpha$ for $MBD_{U}$ always concludes that $MBD(\hat{\bm{Y}})$ and $MBD_{U, \alpha}(\hat{\bm{Y}})$ do not significantly differ for any considered $\alpha$. Therefore, for all the considered models, when functional data are densely observed there is no need to use $MBD_U$ instead of $MBD$\footnote{As a consequence, in the figures in the appendix there are no points for the scenarios ``model 2-setting 1'', ``model 3-setting 1'' and ``model 4-setting 1''.}. Recall that such a decision is taken automatically and in a data-driven way using the proposed procedure to set $\alpha^{*}$. 

\item For models 2, 3 and 4, both $MBD(\bm{\hat{Y}})$ and $MBD_{U, \alpha^{*}}(\bm{\hat{Y}})$ have a stronger relationship with the benchmark $MBD(\bm{Y})$ under setting 2 than under settings 3 or 4. Similar results were observed for model 1.

\item As for model 1, the overall performances of $MBD_{U, \alpha^{*}}(\bm{\hat{Y}})$ are always better than the ones of $MBD(\bm{\hat{Y}})$ for all the new models and under settings 2, 3 and 4. If we consider again $\Delta_{\rho}$, it ranges from 3.77\% (``model 2-setting 4'') to 6.06\% (``model 3-setting 3'').

\item Finally, the pattern of higher $\alpha^{*}$ for more sparsely observed functional data is also observed under models 2, 3 and 4.
\end{itemize}

The simulation study presented in this section provides empirical evidence in favor of using $MBD_{U}$ instead of the standard approach, specially with sparsely observed functional data. In the next section we consider a real functional data set to gather additional information about $MBD_U$.

\section{Real data application: medfly data set}
\label{sec:medfly}
Comparing $MBD$ and $MBD_{U}$ using a real sparse functional data set is a hard task since a benchmark distribution of depth values is missing. Therefore, our strategy consists on considering a real functional data set that has been observed sufficiently densely and might be affected by measurement error, say $\bm{\tilde{Y}}$, and induce sparsity on $\bm{\tilde{Y}}$. We use $MBD(\bm{\tilde{Y}})$ as the benchmark true depth to evaluate $MBD(\bm{\hat{Y}})$ and $MBD_{U;\alpha^{*}}(\bm{\hat{Y}})$, which are computed after inducing sparsity and applying $IEV$ to estimate the curves and obtain the confidence intervals. 

\begin{figure}[!htbp]
\centering
\includegraphics[width=0.66\textwidth]{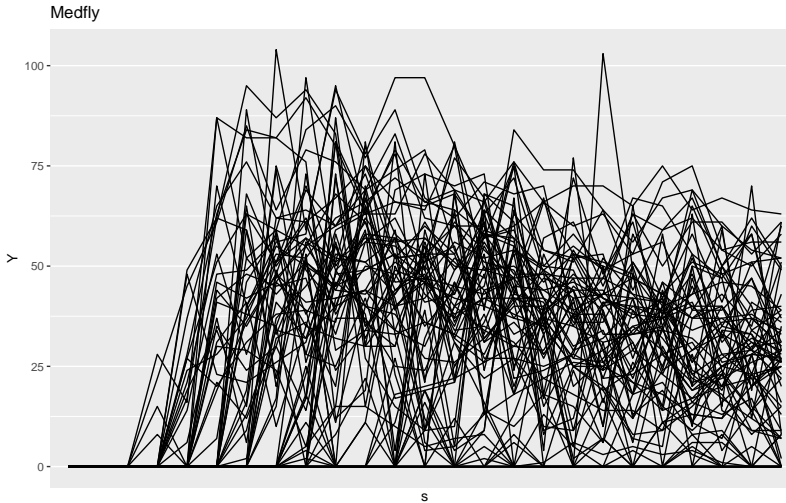}
\captionsetup{width=0.75\textwidth}
\caption{Medfly data set (10\% of the observations).}
\label{fig:f09}
\end{figure}

As $\bm{\tilde{Y}}$ we consider the medfly data set used in \cite{carey1998relationship}. The data set consists on individual egg-laying counts during the first 25 days of lives for 789 female medflies (Mediterranean fruit flies, Ceratitis capitata) at the mass rearing facility in Metapa, Mexico. The medflies egg-laying trajectories are observed at a fix grid of days, $\bm{s}_{g} = \left\{1, \ldots, 25\right\}$. In Figure \ref{fig:f09} we represent the medfly data set (only a 10\% random subsample is represented for graphical reasons).

We induce sparsity on the medfly data set using three different settings. In this case we focus on highly sparse settings:

\begin{itemize}
\item Each curve is observed at $J_{i} = 5$ random points from $\bm{s}_{g}$.

\item Each curve is observed at $J_{i}$ random points from $\bm{s}_{g}$, and $J_{i}$ are i.i.d. from a discrete uniform distribution on the set $\left\{2, 3, 4, 5\right\}$.

\item Each curve is observed at $J_{i} = 2$ random points from $\bm{s}_{g}$.
\end{itemize}

\begin{figure}[!htbp]
\centering
\includegraphics[width=0.66\textwidth]{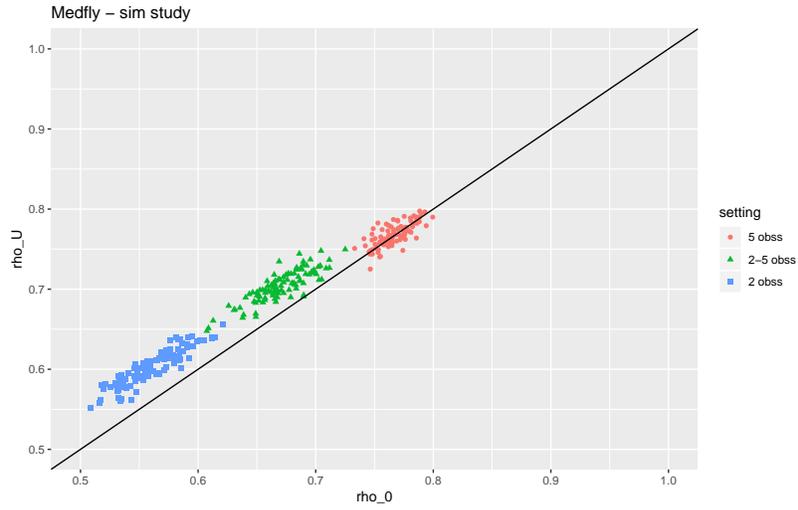}
\captionsetup{width=0.75\textwidth}
\caption{Medfly data set: scatter plot of $(\rho, \rho_{U})$ under different settings.}
\label{fig:f10}
\end{figure}

As in the simulation study, for each setting we simulate 100 sparse data sets. The main difference with respect to the simulation study of Section \ref{sec:sstudy} is that we have a unique sample to which we induce sparsity. Moreover, we observe $\bm{\tilde{Y}}$, i.e., a functional data set that has been possibly measured with error. Therefore $\bm{\tilde{Y}}$ is our benchmark data set. For each simulated sparse data set, after using $IEV$\footnote{We constrain its estimates and confidence intervals to take non-negative values. We use this constrain to be coherent with the nature of the real problem and avoid negative egg-laying counts.}, we compute $MBD(\bm{\hat{Y}})$ and $MBD_{U;\alpha^{*}}(\bm{\hat{Y}})$, and their correlations with the new benchmark $MBD(\bm{\tilde{Y}})$, i.e.,  $\rho_{0} = \rho(MBD(\bm{\tilde{Y}}), MBD(\hat{\bm{Y}}))$ and $\rho_{U} = \rho(MBD(\bm{\tilde{Y}}), MBD_{U, \alpha^{*}}(\hat{\bm{Y}}))$. In Figure \ref{fig:f10} we report all the pairs $(\rho_{0}, \rho_{U})$, whereas in Table \ref{tab:t03} we report the median values of $\rho_{0}, \rho_{U}, \Delta_{\rho}$ and $\alpha^{*}$.

\begin{table}[!htbp]
\captionsetup{width=0.75\textwidth}
\centering
\scalebox{1}{
\begin{tabular}{c|cccc}
\hline
& \multicolumn{4}{c}{median values}\\
\hline
& $\rho_{0}$ & $\rho_{U}$ & $\Delta_{\rho}$ & $\alpha^{*}$\\
\hline
$J_{i} = 5$ & 0.77 & 0.77 & 0.07\% & 0.14\\
$J_{i} \in \left\{2, 3, 4, 5\right\}$ & 0.67 & 0.70 & 5.26\% & 0.35\\
$J_{i} = 2$ & 0.56 & 0.60 & 7.53\% & 0.42\\
\hline
\end{tabular}}
\caption{Medfly data set: median values of $\rho_{0}$, $\rho_{U}$, $\Delta_{\rho}$ and $\alpha^{*}$ under different settings.}
\label{tab:t03}
\end{table}

Based on the results shown in Figure \ref{fig:f10} and Table \ref{tab:t03} we can conclude that the settings with more sparsity show a lower correlation between both $MBD(\bm{\hat{Y}})$ and $MBD_{U;\alpha^{*}}(\bm{\hat{Y}})$, and the benchmark. Under the least sparse setting ($J_{i} = 5$), the performances of $MBD$ and $MBD_{U}$ are comparable, but in the remaining two settings we observe consistent improvements by using $MBD_{U}$. Note that in Figure \ref{fig:f10} the points in the two very sparse scenarios are always above the diagonal line indicating a higher correlation between $MBD_U$ and the benchmark than $MBD$ and the benchmark. With respect to $\alpha^{*}$, we observe a pattern that we have already highlighted in our simulation study: the proposed procedure selects lower confidence levels in relatively more sparse scenarios in a data-driven way. 

Therefore, in very sparse settings our proposal, $MBD_{U;\alpha^{*}}(\bm{\hat{Y}})$, has consistently stronger association to the ``true depth'' $MBD(\bm{\tilde{Y}})$. This result is coherent with the ones found using simulated data, and support the need of taking into account uncertainty when applying depth to sparse functional data.

\section{Conclusions}
In this paper we have introduced a new approach for calculating depth values when functional data sets are sparsely observed. The standard approach consists of using any method for estimating the curves in a fine common grid, possibly borrowing information from the other curves in the sample, and applying a depth function to the estimated curves. We propose to take into account that the curves are estimated with uncertainty and incorporate this key aspect in the calculation of depth values. In particular, we use the $IEV$ method to estimate the curves and their confidence intervals. Moreover, we propose $MBD_{U}$, a new functional depth based on $MBD$ that takes into account both the curve estimates and confidence intervals. Therefore, $MBD_{U}$ incorporates the uncertainty behind the estimation step required for sparse functional data. Finally, with both simulated and real data sets, we have shown the benefits of using $MBD_{U}$ instead of $MBD$ when computing the depth of sparse functional data. In particular, we have shown that $MBD_{U}$ provides a ranking that is closer to the true underlying ranking. 

\newpage

\appendix
\section{The iterated expectation and variance method}
\label{sec:sparse}
$IEV$ assumes that: (1) the underlying curves $Y_{i}(s)$, $1 \leq i \leq n$, are defined in compact interval, $s \in I=[a,b] \subset \mathbb{R}$ and (2) are realizations of a random function with mean $\mu(s) = E\left[Y_{i}(s)\right]$ and covariance operator defined as $\Sigma^{\bm{Y}}(s, s') = Cov\left(Y_{i}(s), Y_{i}(s')\right)$. Based on the spectral decomposition of $\Sigma^{\bm{Y}}(s, s')$, a Karhunen-Lo\`eve expansion for $Y_{i}(s)$ can be defined as $Y_{i}(s) = \mu(s) + \sum_{k = 1}^{\infty} \xi_{ik}\phi_{k}(s)$, where $\bm{\phi}(s) = \left\{\phi_{k}(s): k \in \mathbb{Z}^{+}\right\}$ are orthonormal eigenfunctions, $\lambda_{1} \geq \lambda_{2} \geq \ldots $ are the corresponding nonincreasing eigenvalues , and $\xi_{ik} = \int_{a}^{b} \left\{Y_{i}(s) - \mu(s)\right\} \phi_{k}(s) ds$ are uncorrelated random variables with mean 0 and variance $\lambda_{k}$. Moreover, $IEV$ assumes that: (3) curves are observed with error, i.e., $\tilde{Y}_{i}(s) = Y_{i}(s) + \epsilon_{i}(s)$, where $\tilde{Y}_{i}(s)$ are the observed curves and $\epsilon_{i}(s) \sim N(0, \sigma^{2})$ is the measurement error; (4) $\tilde{Y}_{i}(s)$ are measured on subject specific finite grids $\bm{s}_{i} = \left\{s_{ij}\right\}_{j=1}^{J_{i}}$ that are often irregularly spaced and/or sparse. Note that $J_{i}$ is the number of observations per curve and it is subject dependent. Under these conditions, \cite{goldsmith2013corrected} propose to estimate $\mu(s)$ using penalized splines fit to the pooled observations under working independence and construct $\tilde{\Sigma}^{\bm{Y}}(s, s')$, a raw estimate of the covariance matrix, using a method of moments approach combined with a smoothing step for its off-diagonal elements. Then, the spectral decomposition of $\tilde{\Sigma}^{\bm{Y}}(s, s')$ allows them to obtain the estimated principal component basis functions $\hat{\bm{\phi}}(s) = \left\{\hat{\phi}_{k}(s): k \in \left\{1, \ldots, \hat{K}\right\}\right\}$ and score variances $\left\{\hat{\lambda}_{k}(s): k \in \left\{1, \ldots, \hat{K}\right\}\right\}$, where $\hat{K}$ is the minimum number of components needed to explain 99\% of the variability in the data. They define the final estimation of the covariance matrix as

\begin{equation}
\label{eq:SigmaHat}
\hat{\Sigma}^{\bm{Y}}(s, s') = 
\sum_{k=1}^{\hat{K}} \hat{\lambda}_{k} \hat{\phi}_{k}(s) \hat{\phi}_{k}(s') = 
\hat{\bm{\phi}}(s) \hat{\Lambda} \hat{\bm{\phi}}^{T}(s'),
\end{equation}

\noindent where $\hat{\Lambda}$ is a diagonal matrix with elements $\hat{\lambda}_{1}(s), \ldots, \hat{\lambda}_{\hat{K}}(s)$. Moreover, they estimate $\sigma^{2}$ as the average difference between the middle 60\% diagonal elements of the raw covariance matrix and $\hat{\Sigma}^{\bm{Y}}(s, s')$. 

Following \cite{yao2005functional}, $IEV$ assumes a mixed model framework to predict scores. Given that

\begin{equation}
\label{eq:mixMod01}
\tilde{Y}_{i}(s_{ij}) = \mu(s_{ij}) + \sum_{k=1}^{K} \xi_{ik}\phi_{k}(s_{ij}) + \epsilon_{i}(s_{ij}),
\end{equation}

\noindent and

\begin{equation}
\label{eq:mixMod02}
\bm{\xi}_{i} \overset{iid}{\sim} N[0, \Lambda],\ \epsilon_{i}(s_{ij}) \overset{iid}{\sim} N[0, \sigma^{2}],
\end{equation}

\noindent where $\bm{\xi}_{i} = \left\{\xi_{ik}: k \in \left\{1, \ldots, K\right\} \right\}$ and $\bm{\epsilon}_{i} = \left\{\epsilon_{i}(s_{i1}), \ldots, \epsilon_{i}(s_{iJ_{i}})\right\}$ are independent, the scores can be estimated using their Best Linear Unbiased Predictions (BLUPs), i.e.,  

\begin{equation}
\label{eq:BLUPS}
\hat{\bm{\xi}}_{\hat{\bm{\theta}}, i} = 
E\left[\bm{\xi}_{i}\left|\right.\tilde{Y}_{i}(\bm{s}_{i}), \hat{\bm{\theta}}\right] =
\left(\hat{\bm{\phi}}^{T}(\bm{s}_i) \hat{\bm{\phi}}(\bm{s}_i) + \hat{\sigma}^2 \hat{\Lambda}^{-1}\right)^{-1} \times 
\hat{\bm{\phi}}^{T}(\bm{s}_{i})\left(\tilde{Y}_{i}(\bm{s}_{i}) - \hat{\mu}(\bm{s}_{i})\right),
\end{equation}

\noindent where $\bm{\theta} = \{\bm{\phi}(s), \mu(s), \Lambda, \sigma^{2}, K\}$ is the collection of unobserved FPC decomposition objects, $\hat{\bm{\theta}}$ is its estimate, and $\tilde{Y}_{i}(\bm{s}_{i}), \hat{\bm{\phi}}(\bm{s}_i)$ and $\hat{\mu}(\bm{s}_{i})$ are the vector of observations for curve $i$, the $J_{i} \times K$ matrix containing the collection of estimated basis functions evaluated at $\bm{s}_{i}$ and the estimated mean function evaluated at $\bm{s}_{i}$, respectively.

Given a particular decomposition, the method proposed by \cite{yao2005functional} provides not only the scores estimates in \eqref{eq:BLUPS},  but also the estimate of $Y_{i}$ over the dense grid $\bm{s}_{g}$ as

\begin{equation}
\label{eq:tildeY}
\hat{Y}_{\hat{\bm{\theta}}, i}(\bm{s}_{g}) = 
E\left[\tilde{Y}_{i}(\bm{s}_{g})\left|\right.\hat{\bm{\xi}}_{\hat{\bm{\theta}}, i}, \hat{\bm{\theta}}\right] =
\hat{\mu}(\bm{s}_{g}) + \hat{\bm{\phi}}(\bm{s}_g)\hat{\bm{\xi}}_{\hat{\bm{\theta}}, i},
\end{equation} 

\noindent where $\bm{s}_{g}$ is often taken as the union of the $\bm{s}_{i}$, while the covariance operator of \eqref{eq:tildeY} is given by

\begin{equation}
\label{eq:covOp}
Var\left[\hat{Y}_{\hat{\bm{\theta}}, i}(\bm{s}_{g}) - Y_{i}(\bm{s}_{g}) \left|\right. \hat{\bm{\theta}}\right] \approx \hat{\bm{\phi}}(\bm{s}_{g}) \left(\frac{1}{\hat{\sigma}^2} \hat{\bm{\phi}}^{T}(\bm{s}_{i}) \hat{\bm{\phi}}(\bm{s}_{i}) + \hat{\Lambda}^{-1}\right)^{-1} \hat{\bm{\phi}}^{T}(\bm{s}_{g}).
\end{equation}

\noindent Using \eqref{eq:tildeY} and \eqref{eq:covOp}, \cite{yao2005functional} derived the $100(1-\alpha)\%$ point-wise confidence intervals for $\hat{Y}_{\hat{\bm{\theta}}, i}(\bm{s}_{g})$ in the following way:

\begin{equation}
\label{eq:pCIs}
\hat{Y}_{\hat{\bm{\theta}}, i}(\bm{s}_{g}) \pm \Phi^{-1}\left(1- \frac{\alpha}{2}\right) \sqrt{\mbox{diag}\left\{Var\left[\hat{Y}_{\hat{\bm{\theta}}, i}(\bm{s}_{g}) - Y_{i}(\bm{s}_{g}) \left|\right. \hat{\bm{\theta}}\right]\right\}}, 
\end{equation} 

\noindent where $\Phi(\cdot)$ is the standard Gaussian cumulative distribution function.

$IEV$ modifies the previous procedure by implementing a bootstrap step to take into account PCA decomposition-based uncertainty. In the $IEV$ method, bootstrap samples are obtained by resampling curves with replacement. Given the $b$th bootstrap sample $\tilde{\bm{Y}}_b$, they derive the bootstrap analogues of \eqref{eq:SigmaHat} and $\hat{\bm{\theta}}$, denoted as $\hat{\Sigma}_{b}^{\bm{Y}}$ and $\hat{\bm{\theta}}_{b}$. Conditioning on $\hat{\bm{\theta}}_{b}$, the bootstrap analogues of \eqref{eq:BLUPS}, \eqref{eq:tildeY} and \eqref{eq:covOp}, i.e., $\hat{\bm{\xi}}_{\hat{\bm{\theta}}_{b}, i}$, $\hat{Y}_{\hat{\bm{\theta}}_{b}, i}(\bm{s}_{g})$ and $Var\left[\hat{Y}_{\hat{\bm{\theta}}_{b}, i}(\bm{s}_{g}) - Y_{i}(\bm{s}_{g}) \left|\right. \hat{\bm{\theta}}_{b}\right]$, for each element of the full data set $\tilde{\bm{Y}}$ can be obtained. Finally, they combine information across bootstrap samples to estimate curves and construct variability estimates. Using the iterated expectation formula,  the estimate of $Y_{i}(s)$ over the dense grid $\bm{s}_{g}$ is given by

\begin{equation}
\label{eq:tildeYIEV}
\hat{Y}_{i}(\bm{s}_{g}) = 
E_{\hat{\bm{\theta}}}\left\{E_{\tilde{Y}_{i}\left|\right.\hat{\bm{\theta}}}\left[\tilde{Y}_{i}(\bm{s}_{g})\left|\right.\hat{\bm{\xi}}_{\hat{\bm{\theta}}, i}, \hat{\bm{\theta}}\right]\right\},
\end{equation} 

\noindent and using the iterated variance formula, the total covariance operator of the estimated curves is given by:

\begin{equation}
\label{eq:covOpIEV}
Var\left[\hat{Y}_{\hat{\bm{\theta}}, i}(\bm{s}_{g}) - Y_{i}(\bm{s}_{g})\right] = 
E_{\hat{\bm{\theta}}}\left[Var_{\tilde{Y}\left|\right.\hat{\bm{\theta}}}\left(\hat{Y}_{\hat{\bm{\theta}}, i}(\bm{s}_{g}) - Y_{i}(\bm{s}_{g})\left|\right.\hat{\bm{\theta}}\right)\right] +
Var_{\hat{\bm{\theta}}}\left[E_{\tilde{Y}\left|\right.\hat{\bm{\theta}}}\left(\hat{Y}_{\hat{\bm{\theta}}, i}(\bm{s}_{g}) - Y_{i}(\bm{s}_{g})\left|\right.\hat{\bm{\theta}}\right)\right].
\end{equation}

\noindent For a confidence level $100(1-\alpha)\%$, \cite{goldsmith2013corrected} obtained (point-wise) confidence intervals for $\hat{Y}_{i}(\bm{s}_{g})$ in the following way:

\begin{equation}
\label{eq:pCIsIEV}
\hat{Y}_{i}(\bm{s}_{g}) \pm \Phi^{-1}\left(1- \frac{\alpha}{2}\right) \sqrt{\mbox{diag}\left\{Var\left[\hat{Y}_{i}(\bm{s}_{g}) - Y_{i}(\bm{s}_{g})\right]\right\}}. 
\end{equation} 

\noindent Note that \eqref{eq:pCIsIEV} is based on $Var\left[\hat{Y}_{i}(\bm{s}_{g}) - Y_{i}(\bm{s}_{g})\right]$, which is approximated by $Var\left[\hat{Y}_{\hat{\bm{\theta}}, i}(\bm{s}_{g}) - Y_{i}(\bm{s}_{g})\right]$. The width of this confidence interval is subject dependent and will vary depending on the sparsity of the specific observation. Given an $\alpha$, those curves observed at fewer points will have a wider confidence interval indicating more uncertainty in the estimation.

\newpage

\section{Simulation study: figures for models 2, 3 and 4}
\begin{figure}[!htbp]
\centering
\includegraphics[width=0.66\textwidth]{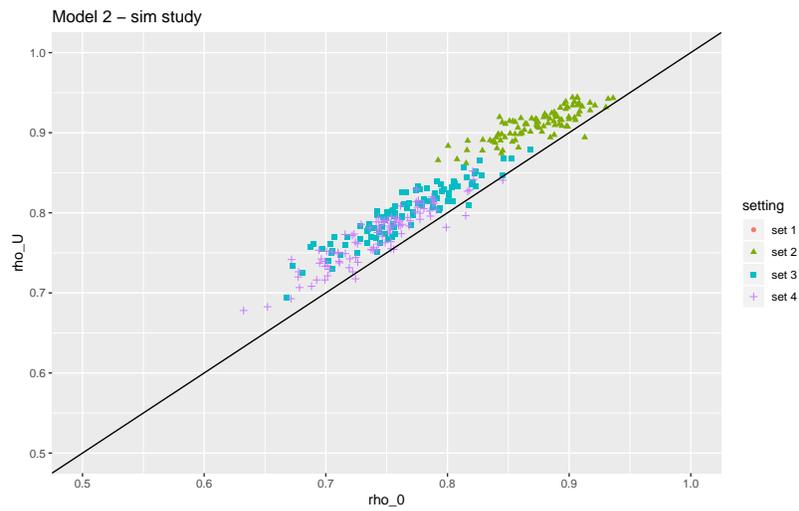}
\captionsetup{width=0.75\textwidth}
\caption{Model 2: scatter plot of $(\rho_{0}, \rho_{U})$ under settings 1-4.}
\label{fig:f06}
\end{figure}

\begin{figure}[!htbp]
\centering
\includegraphics[width=0.66\textwidth]{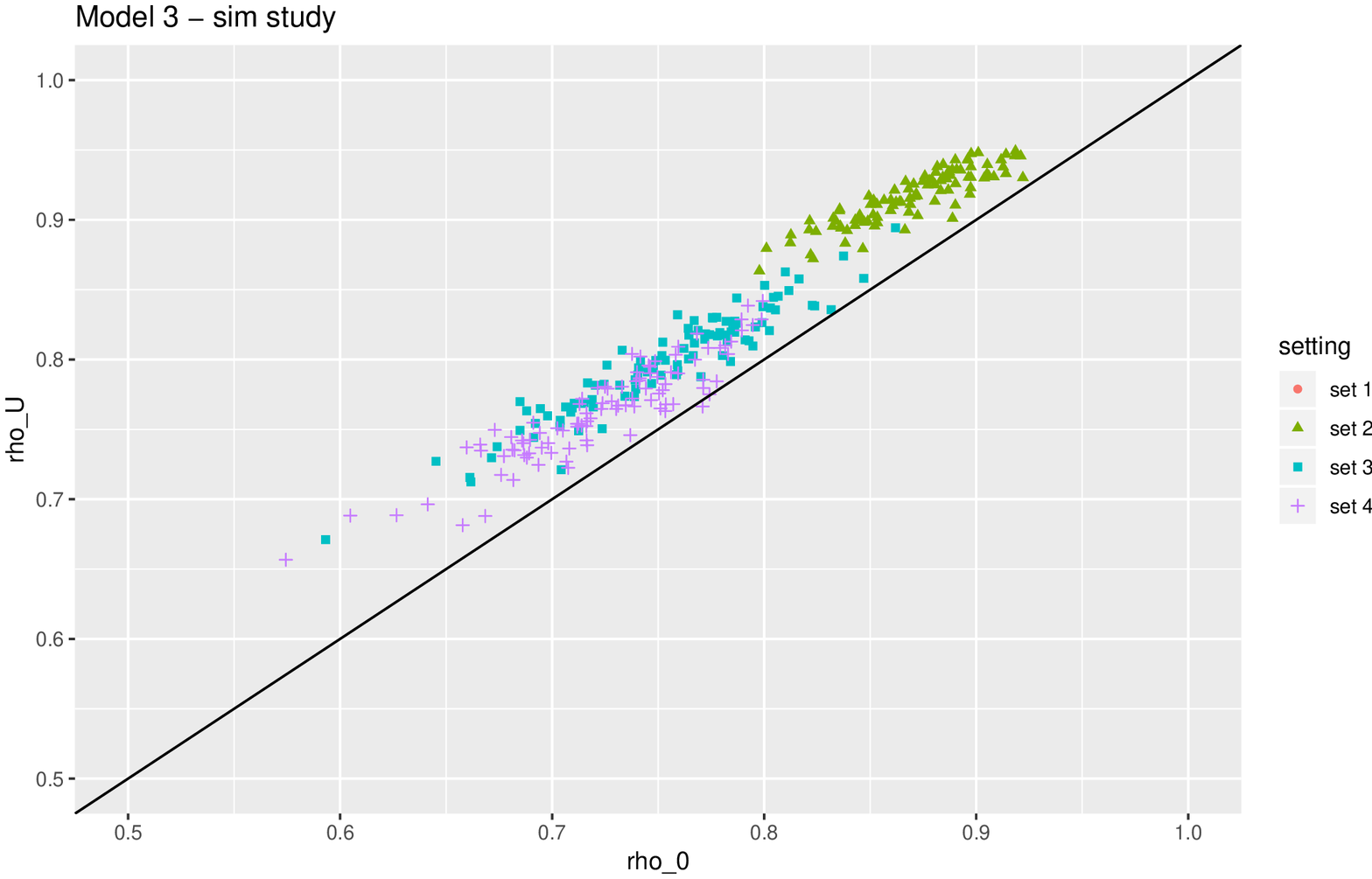}
\captionsetup{width=0.75\textwidth}
\caption{Model 3: scatter plot of $(\rho_{0}, \rho_{U})$ under settings 1-4.}
\label{fig:f07}
\end{figure}

\begin{figure}[!htbp]
\centering
\includegraphics[width=0.66\textwidth]{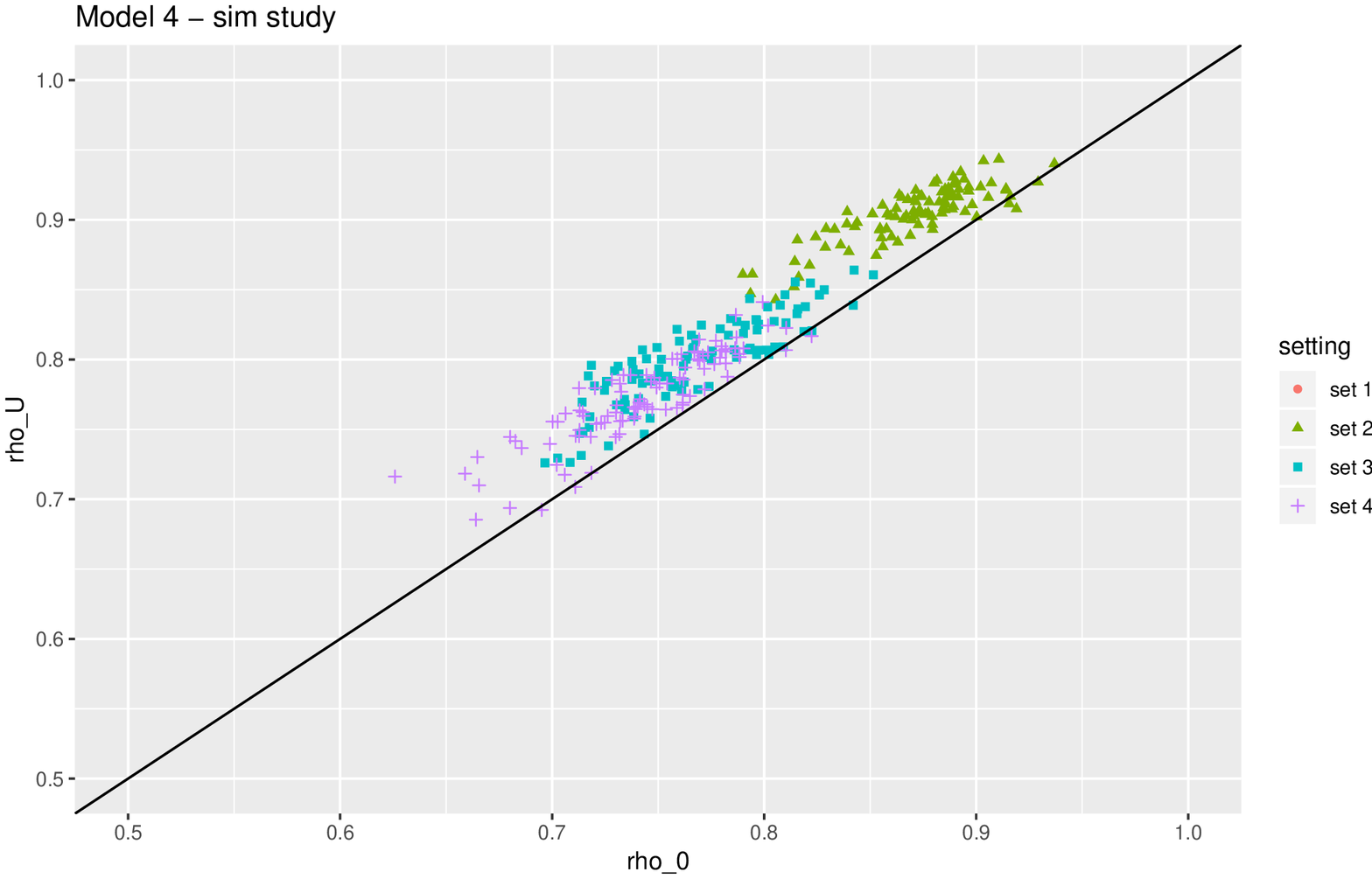}
\captionsetup{width=0.75\textwidth}
\caption{Model 4: scatter plot of $(\rho_{0}, \rho_{U})$ under settings 1-4.}
\label{fig:f08}
\end{figure}

\newpage

\phantom{white page}

\newpage

\bibliographystyle{Chicago}
\bibliography{Bibliography-MM-MC2}

\begin{thebibliography}{}

\bibitem[\protect\citeauthoryear{Arribas-Gil and Romo}{Arribas-Gil and
  Romo}{2014}]{arribas2014shape}
Arribas-Gil, A. and J.~Romo (2014).
\newblock Shape outlier detection and visualization for functional data: the
  outliergram.
\newblock {\em Biostatistics\/}~{\em 15\/}(4), 603--619.

\bibitem[\protect\citeauthoryear{Azcorra, Chiroque, Cuevas, Anta, Laniado,
  Lillo, Romo, and Sguera}{Azcorra et~al.}{2018}]{azcorra2018unsupervised}
Azcorra, A., L.~F. Chiroque, R.~Cuevas, A.~F. Anta, H.~Laniado, R.~E. Lillo,
  J.~Romo, and C.~Sguera (2018).
\newblock Unsupervised scalable statistical method for identifying influential
  users in online social networks.
\newblock {\em Scientific reports\/}~{\em 8\/}(1), 6955.

\bibitem[\protect\citeauthoryear{Carey, Liedo, M{\"u}ller, Wang, and
  Chiou}{Carey et~al.}{1998}]{carey1998relationship}
Carey, J.~R., P.~Liedo, H.-G. M{\"u}ller, J.-L. Wang, and J.-M. Chiou (1998).
\newblock Relationship of age patterns of fecundity to mortality, longevity,
  and lifetime reproduction in a large cohort of mediterranean fruit fly
  females.
\newblock {\em The Journals of Gerontology Series A: Biological Sciences and
  Medical Sciences\/}~{\em 53\/}(4), B245--B251.

\bibitem[\protect\citeauthoryear{Chakraborty and Chaudhuri}{Chakraborty and
  Chaudhuri}{2014}]{chakraborty2014data}
Chakraborty, A. and P.~Chaudhuri (2014).
\newblock On data depth in infinite dimensional spaces.
\newblock {\em Annals of the Institute of Statistical Mathematics\/}~{\em
  66\/}(2), 303--324.

\bibitem[\protect\citeauthoryear{Chaudhuri}{Chaudhuri}{1996}]{chaudhuri1996geometric}
Chaudhuri, P. (1996).
\newblock On a geometric notion of quantiles for multivariate data.
\newblock {\em Journal of the American Statistical Association\/}~{\em
  91\/}(434), 862--872.

\bibitem[\protect\citeauthoryear{Cuesta-Albertos, Febrero-Bande, and de~la
  Fuente}{Cuesta-Albertos et~al.}{2017}]{cuesta2017hbox}
Cuesta-Albertos, J.~A., M.~Febrero-Bande, and M.~O. de~la Fuente (2017).
\newblock The ddg-classifier in the functional setting.
\newblock {\em Test\/}~{\em 26\/}(1), 119--142.

\bibitem[\protect\citeauthoryear{Cuesta-Albertos and
  Nieto-Reyes}{Cuesta-Albertos and Nieto-Reyes}{2008}]{cuesta2008random}
Cuesta-Albertos, J.~A. and A.~Nieto-Reyes (2008).
\newblock The random tukey depth.
\newblock {\em Computational Statistics \& Data Analysis\/}~{\em 52\/}(11),
  4979--4988.

\bibitem[\protect\citeauthoryear{Cuevas, Febrero, and Fraiman}{Cuevas
  et~al.}{2007}]{cuevas2007robust}
Cuevas, A., M.~Febrero, and R.~Fraiman (2007).
\newblock Robust estimation and classification for functional data via
  projection-based depth notions.
\newblock {\em Computational Statistics\/}~{\em 22\/}(3), 481--496.

\bibitem[\protect\citeauthoryear{Dai and Genton}{Dai and
  Genton}{2017}]{dai2017outlyingness}
Dai, W. and M.~G. Genton (2017).
\newblock An outlyingness matrix for multivariate functional data
  classification.
\newblock {\em arXiv preprint arXiv:1704.02568\/}.

\bibitem[\protect\citeauthoryear{Flores, Lillo, and Romo}{Flores
  et~al.}{2018}]{flores2018homogeneity}
Flores, R., R.~Lillo, and J.~Romo (2018).
\newblock Homogeneity test for functional data.
\newblock {\em Journal of Applied Statistics\/}~{\em 45\/}(5), 868--883.

\bibitem[\protect\citeauthoryear{Fraiman and Muniz}{Fraiman and
  Muniz}{2001}]{fraiman2001trimmed}
Fraiman, R. and G.~Muniz (2001).
\newblock Trimmed means for functional data.
\newblock {\em Test\/}~{\em 10\/}(2), 419--440.

\bibitem[\protect\citeauthoryear{Gervini}{Gervini}{2012}]{gervini2012outlier}
Gervini, D. (2012).
\newblock Outlier detection and trimmed estimation for general functional data.
\newblock {\em Statistica Sinica\/}, 1639--1660.

\bibitem[\protect\citeauthoryear{Gijbels and Nagy}{Gijbels and
  Nagy}{2017}]{gijbels2017general}
Gijbels, I. and S.~Nagy (2017).
\newblock On a general definition of depth for functional data.
\newblock {\em Statistical Science\/}~{\em 32\/}(4), 630--639.

\bibitem[\protect\citeauthoryear{Goldsmith, Greven, and Crainiceanu}{Goldsmith
  et~al.}{2013}]{goldsmith2013corrected}
Goldsmith, J., S.~Greven, and C.~M. Crainiceanu (2013).
\newblock Corrected confidence bands for functional data using principal
  components.
\newblock {\em Biometrics\/}~{\em 69\/}(1), 41--51.

\bibitem[\protect\citeauthoryear{Hubert, Rousseeuw, and Segaert}{Hubert
  et~al.}{2015}]{hubert2015multivariate}
Hubert, M., P.~J. Rousseeuw, and P.~Segaert (2015).
\newblock Multivariate functional outlier detection.
\newblock {\em Statistical Methods \& Applications\/}~{\em 24\/}(2), 177--202.

\bibitem[\protect\citeauthoryear{J{\"o}rnsten}{J{\"o}rnsten}{2004}]{jornsten2004clustering}
J{\"o}rnsten, R. (2004).
\newblock Clustering and classification based on the l1 data depth.
\newblock {\em Journal of Multivariate Analysis\/}~{\em 90\/}(1), 67--89.

\bibitem[\protect\citeauthoryear{Koshevoy, Mosler, et~al.}{Koshevoy
  et~al.}{1997}]{koshevoy1997zonoid}
Koshevoy, G., K.~Mosler, et~al. (1997).
\newblock Zonoid trimming for multivariate distributions.
\newblock {\em The Annals of Statistics\/}~{\em 25\/}(5), 1998--2017.

\bibitem[\protect\citeauthoryear{Li, Cuesta-Albertos, and Liu}{Li
  et~al.}{2012}]{li2012dd}
Li, J., J.~A. Cuesta-Albertos, and R.~Y. Liu (2012).
\newblock Dd-classifier: Nonparametric classification procedure based on
  dd-plot.
\newblock {\em Journal of the American Statistical Association\/}~{\em
  107\/}(498), 737--753.

\bibitem[\protect\citeauthoryear{Liu et~al.}{Liu et~al.}{1990}]{liu1990notion}
Liu, R.~Y. et~al. (1990).
\newblock On a notion of data depth based on random simplices.
\newblock {\em The Annals of Statistics\/}~{\em 18\/}(1), 405--414.

\bibitem[\protect\citeauthoryear{Liu, Parelius, Singh, et~al.}{Liu
  et~al.}{1999}]{liu1999multivariate}
Liu, R.~Y., J.~M. Parelius, K.~Singh, et~al. (1999).
\newblock Multivariate analysis by data depth: descriptive statistics, graphics
  and inference,(with discussion and a rejoinder by liu and singh).
\newblock {\em The annals of statistics\/}~{\em 27\/}(3), 783--858.

\bibitem[\protect\citeauthoryear{Liu and Singh}{Liu and
  Singh}{1993}]{liu1993quality}
Liu, R.~Y. and K.~Singh (1993).
\newblock A quality index based on data depth and multivariate rank tests.
\newblock {\em Journal of the American Statistical Association\/}~{\em
  88\/}(421), 252--260.

\bibitem[\protect\citeauthoryear{L{\'o}pez-Pintado and
  Jornsten}{L{\'o}pez-Pintado and Jornsten}{2007}]{lopez2007functional}
L{\'o}pez-Pintado, S. and R.~Jornsten (2007).
\newblock Functional analysis via extensions of the band depth.
\newblock {\em Lecture Notes-Monograph Series\/}, 103--120.

\bibitem[\protect\citeauthoryear{L{\'o}pez-Pintado and Romo}{L{\'o}pez-Pintado
  and Romo}{2007}]{lopez2007depth}
L{\'o}pez-Pintado, S. and J.~Romo (2007).
\newblock Depth-based inference for functional data.
\newblock {\em Computational Statistics \& Data Analysis\/}~{\em 51\/}(10),
  4957--4968.

\bibitem[\protect\citeauthoryear{L{\'o}pez-Pintado and Romo}{L{\'o}pez-Pintado
  and Romo}{2009}]{lopez2009concept}
L{\'o}pez-Pintado, S. and J.~Romo (2009).
\newblock On the concept of depth for functional data.
\newblock {\em Journal of the American Statistical Association\/}~{\em
  104\/}(486), 718--734.

\bibitem[\protect\citeauthoryear{L{\'o}pez-Pintado and Romo}{L{\'o}pez-Pintado
  and Romo}{2011}]{lopez2011half}
L{\'o}pez-Pintado, S. and J.~Romo (2011).
\newblock A half-region depth for functional data.
\newblock {\em Computational Statistics \& Data Analysis\/}~{\em 55\/}(4),
  1679--1695.

\bibitem[\protect\citeauthoryear{L{\'o}pez-Pintado and Wei}{L{\'o}pez-Pintado
  and Wei}{2011}]{lopez2011depth}
L{\'o}pez-Pintado, S. and Y.~Wei (2011).
\newblock Depth for sparse functional data.
\newblock In {\em Recent advances in functional data analysis and related
  topics}, pp.\  209--212. Springer.

\bibitem[\protect\citeauthoryear{L{\'o}pez-Pintado and
  Wrobel}{L{\'o}pez-Pintado and Wrobel}{2017}]{lopez2017robust}
L{\'o}pez-Pintado, S. and J.~Wrobel (2017).
\newblock Robust non-parametric tests for imaging data based on data depth.
\newblock {\em Stat\/}~{\em 6\/}(1), 405--419.

\bibitem[\protect\citeauthoryear{Mahalanobis}{Mahalanobis}{1936}]{mahalanobis1936generalized}
Mahalanobis, P.~C. (1936).
\newblock On the generalized distance in statistics.
\newblock National Institute of Science of India.

\bibitem[\protect\citeauthoryear{Mosler and Polyakova}{Mosler and
  Polyakova}{2012}]{mosler2012general}
Mosler, K. and Y.~Polyakova (2012).
\newblock General notions of depth for functional data.
\newblock {\em arXiv preprint arXiv:1208.1981\/}.

\bibitem[\protect\citeauthoryear{Narisetty and Nair}{Narisetty and
  Nair}{2016}]{narisetty2016extremal}
Narisetty, N.~N. and V.~N. Nair (2016).
\newblock Extremal depth for functional data and applications.
\newblock {\em Journal of the American Statistical Association\/}~{\em
  111\/}(516), 1705--1714.

\bibitem[\protect\citeauthoryear{Nieto-Reyes and Battey}{Nieto-Reyes and
  Battey}{2016}]{nieto2016topologically}
Nieto-Reyes, A. and H.~Battey (2016).
\newblock A topologically valid definition of depth for functional data.
\newblock {\em Statistical Science\/}, 61--79.

\bibitem[\protect\citeauthoryear{Oja}{Oja}{1983}]{oja1983descriptive}
Oja, H. (1983).
\newblock Descriptive statistics for multivariate distributions.
\newblock {\em Statistics \& Probability Letters\/}~{\em 1\/}(6), 327--332.

\bibitem[\protect\citeauthoryear{Rousseeuw and Hubert}{Rousseeuw and
  Hubert}{1999}]{rousseeuw1999regression}
Rousseeuw, P.~J. and M.~Hubert (1999).
\newblock Regression depth.
\newblock {\em Journal of the American Statistical Association\/}~{\em
  94\/}(446), 388--402.

\bibitem[\protect\citeauthoryear{Sguera, Galeano, and Lillo}{Sguera
  et~al.}{2014}]{sguera2014spatial}
Sguera, C., P.~Galeano, and R.~Lillo (2014).
\newblock Spatial depth-based classification for functional data.
\newblock {\em Test\/}~{\em 23\/}(4), 725--750.

\bibitem[\protect\citeauthoryear{Sguera, Galeano, and Lillo}{Sguera
  et~al.}{2016}]{sguera2016functional}
Sguera, C., P.~Galeano, and R.~E. Lillo (2016).
\newblock Functional outlier detection by a local depth with application to no
  x levels.
\newblock {\em Stochastic environmental research and risk assessment\/}~{\em
  30\/}(4), 1115--1130.

\bibitem[\protect\citeauthoryear{Sun and Genton}{Sun and
  Genton}{2011}]{sun2011functional}
Sun, Y. and M.~G. Genton (2011).
\newblock Functional boxplots.
\newblock {\em Journal of Computational and Graphical Statistics\/}~{\em
  20\/}(2), 316--334.

\bibitem[\protect\citeauthoryear{Sun and Genton}{Sun and
  Genton}{2012}]{sun2012functional}
Sun, Y. and M.~G. Genton (2012).
\newblock Functional median polish.
\newblock {\em Journal of agricultural, biological, and environmental
  statistics\/}~{\em 17\/}(3), 354--376.

\bibitem[\protect\citeauthoryear{Tukey}{Tukey}{1975}]{tukey1975mathematics}
Tukey, J.~W. (1975).
\newblock Mathematics and the picturing of data.
\newblock In {\em Proceedings of the International Congress of Mathematicians,
  Vancouver, 1975}, Volume~2, pp.\  523--531.

\bibitem[\protect\citeauthoryear{Vardi and Zhang}{Vardi and
  Zhang}{2000}]{vardi2000multivariate}
Vardi, Y. and C.-H. Zhang (2000).
\newblock The multivariate l1-median and associated data depth.
\newblock {\em Proceedings of the National Academy of Sciences\/}~{\em
  97\/}(4), 1423--1426.

\bibitem[\protect\citeauthoryear{Yao, M{\"u}ller, and Wang}{Yao
  et~al.}{2005}]{yao2005functional}
Yao, F., H.-G. M{\"u}ller, and J.-L. Wang (2005).
\newblock Functional data analysis for sparse longitudinal data.
\newblock {\em Journal of the American Statistical Association\/}~{\em
  100\/}(470), 577--590.

\bibitem[\protect\citeauthoryear{Zhang and Wang}{Zhang and
  Wang}{2016}]{zhang2016sparse}
Zhang, X. and J.-L. Wang (2016).
\newblock From sparse to dense functional data and beyond.
\newblock {\em The Annals of Statistics\/}~{\em 44\/}(5), 2281--2321.

\bibitem[\protect\citeauthoryear{Zuo}{Zuo}{2003}]{zuo2003projection}
Zuo, Y. (2003).
\newblock Projection-based depth functions and associated medians.
\newblock {\em The Annals of Statistics\/}~{\em 31\/}(5), 1460--1490.

\bibitem[\protect\citeauthoryear{Zuo and Serfling}{Zuo and
  Serfling}{2000}]{zuo2000general}
Zuo, Y. and R.~Serfling (2000).
\newblock General notions of statistical depth function.
\newblock {\em Annals of statistics\/}, 461--482.

\end{thebibliography}

\end{document}